\documentclass[12pt,preprint]{aastex}

\def\eg{\emph{e.g. }}
\def\et{\emph{et al.}}
\def\ie{\emph{i.e. }}

\def\apj{\emph{ApJ}}
\def\apjl{\emph{ApJ Letters}}
\def\apjs{\emph{ApJ Supp.}}

\def\aa{\emph{A \& A}}

\def\nat{\emph{Nature}}

\def\x{$\times \, \,$}
\def\about{$\sim$}

\shorttitle{The Energy Spectrum of Mrk 421}
\shortauthors{Carson, \et}

\begin{document}

\title{The Energy Spectrum of the Blazar Markarian 421 \\
Above 130 GeV}

\author{J.E. Carson\altaffilmark{1,a}}

\author{J. Kildea\altaffilmark{2,b}}

\author{R. A. Ong\altaffilmark{1}}

\author{J. Ball\altaffilmark{1}}

\author{D. A. Bramel\altaffilmark{3,c}}

\author{C. E. Covault\altaffilmark{4}}

\author{D. Driscoll\altaffilmark{4}}

\author{P. Fortin\altaffilmark{5}}

\author{D. M. Gingrich\altaffilmark{6,7}} 

\author{D. S. Hanna\altaffilmark{2}} 

\author{T. Lindner\altaffilmark{2,d}}

\author{C. Mueller\altaffilmark{2}}

\author{A. Jarvis\altaffilmark{1}}

\author{R. Mukherjee\altaffilmark{5}}

\author{K. Ragan\altaffilmark{2}}

\author{R. A. Scalzo\altaffilmark{8,e}}

\author{D. A. Williams\altaffilmark{9}}

\author{J. Zweerink\altaffilmark{1}}

\altaffiltext{1}{Department of Physics and Astronomy, University of California, Los Angeles, Los Angeles, CA 90095}
\altaffiltext{a}{Current address: Stanford Linear Accelerator Center, MS 29, Menlo Park, CA 94025}
\altaffiltext{2}{Department of Physics, McGill University, Montreal, QC H3A 2T8, Canada}
\altaffiltext{b}{Current address: Fred Lawrence Whipple Observatory, Harvard-Smithsonian Center for Astrophysics, Amado, AZ 85645}
\altaffiltext{3}{Department of Physics, Columbia University, New York, NY 10027}
\altaffiltext{c}{Current address: Interactive Brokers, Greenwich, CT}
\altaffiltext{4}{Department of Physics, Case Western Reserve University, Cleveland, OH 44106}
\altaffiltext{5}{Department of Physics and Astronomy, Barnard College, Columbia University, New York, NY 10027}
\altaffiltext{6}{Department of Physics, University of Alberta, Edmonton, AB T6G 2G7 Canada}
\altaffiltext{7}{Also at TRIUMF, Vancouver, BC V6T 2A3 Canada}
\altaffiltext{d}{Current address: Department of Physics and Astronomy, University of British Columbia, Vancouver, BC V6T 1Z1, Canada}
\altaffiltext{8}{Department of Physics, University of Chicago, Chicago, IL 60637}
\altaffiltext{e}{Current address: Lawrence Berkeley National Laboratory, Berkeley, CA 94720}
\altaffiltext{9}{Santa Cruz Institute for Particle Physics, University of California, Santa Cruz, Santa Cruz, CA 95064} 

\begin{abstract}

Markarian 421 (Mrk 421) was the first blazar detected at gamma-ray
energies above 300 GeV, and it remains one of only twelve TeV blazars
detected to date. TeV gamma-ray measurements of its flaring activity
and spectral variability have placed constraints on models of the
high-energy emission from blazars. However, observations between 50
and 300 GeV are rare, and the high-energy peak of the spectral energy
distribution (SED), predicted to be in this range, has never been
directly detected. We present a detection of Mrk 421 above 100 GeV as
made by the Solar Tower Atmospheric Cherenkov Effect Experiment
(STACEE) during a multiwavelength campaign in early 2004. STACEE is a
ground-based atmospheric Cherenkov telescope using the wavefront
sampling technique to detect gamma rays at lower energies than
achieved by most imaging Cherenkov telescopes. We also outline a
method for reconstructing gamma-ray energies using a solar heliostat
telescope. This technique was applied to the 2004 data, and we present
the differential energy spectrum of Mrk 421 above 130 GeV. Assuming a
differential photon flux $\frac{dN}{dE} \propto E^{-\alpha}$, we
measure a spectral index $\alpha = 2.1 \pm 0.2_{stat} \, {^{+0.2}
_{-0.1}}_{sys}$. Finally, we discuss the STACEE spectrum in the
context of the multiwavelength results from the same epoch.

\end{abstract}


\section{\label{intro}Introduction}

The X-ray-bright, nearby (z = 0.03) blazar Markarian 421 was the first
extragalactic object detected at very high energies (VHE, $E > 100$
GeV) \citep{punch92} and remains one of a dozen VHE detections of
blazars to date. Since its discovery, it has produced several major
flares, providing many opportunities for multiwavelength campaigns
(\eg \citet{gaidos96,krennrich99,piron01,aharonian02}). During these
periods of days to weeks, it outshines the Crab Nebula by factors of
$2 - 10$.

The gamma-ray spectrum of Mrk 421 above 300 GeV has been measured
several times at different flux levels, and the VHE spectral index
$\alpha$ is known to vary significantly (differential flux
$\frac{dN}{dE} \propto E^{-\alpha}$). \citet{krennrich02} and
\citet{aharonian02} made the important observation that the spectral
index and the absolute flux level are correlated, suggesting that the
high-energy peak of the spectral energy distribution (SED) shifts to
higher energies during periods of heightened activity. Also, above a
few TeV, a curved deviation from a constant power law spectrum has
been detected for Mrk 421 \citep{krennrich01}.

The simultaneous radio to VHE SED of Mrk 421 has been successfully
described in terms of synchrotron self-Compton (SSC) models
\citep{maraschi99,fossati00,krawczynski01,blazejowski05}. In these
models, synchrotron photons from relativistic electrons in the inner
regions of the magnetized jet gain energy via inverse-Compton
scattering off the same electron population that radiated them. The
synchrotron spectrum peaks in the X-ray band, while the
inverse-Compton photons are radiated at GeV and TeV energies. Within
this framework, the X-ray and TeV data have constrained the magnetic
field $B$ and the size $R$ of the gamma-ray emitting region to within
factors of a few: $B \sim 0.1$ Gauss and $R \sim 10^{16}$ cm. The
measurements are also beginning to constrain the Doppler factor of the
bulk material in the jet, $\delta_{jet} \sim 10-50$, and the shape of
the electron energy distribution.

Although well-studied at other energies, the flux of Mrk 421 between
100 and 300 GeV has only been measured three times
\citep{boone02,piron03,albert06}, and only two measurements have been
made of its spectrum around 100 GeV
\citep{piron03,albert06}. Measurements in this waveband are of
particular scientific interest, because most SSC models predict that
the high-energy SED will peak around 100 GeV. Testing this prediction
can place valuable constraints on the various electron emission
models, especially in describing the energy distribution of the
relativistic electrons.

We report the detection of Mrk 421 above 100 GeV using the Solar Tower
Atmospheric Cherenkov Effect Experiment (STACEE, \citet{gingrich05})
and present a measurement of the integral flux and differential energy
spectrum between 130 GeV and 2 TeV. Section \ref{sec: observations}
briefly describes the STACEE experiment itself and the Mrk 421 data
set. Section \ref{sec: analysis} presents the basic data analysis, and
Section \ref{sec: int results} presents the integral flux
results. Section \ref{sec: spec results} describes the spectral
analysis and results, and Section \ref{sec: discussion} discusses the
scientific implications of the results.

\section{\label{sec: observations}STACEE Observations}

\begin{figure}
\epsscale{1.0}
\plotone{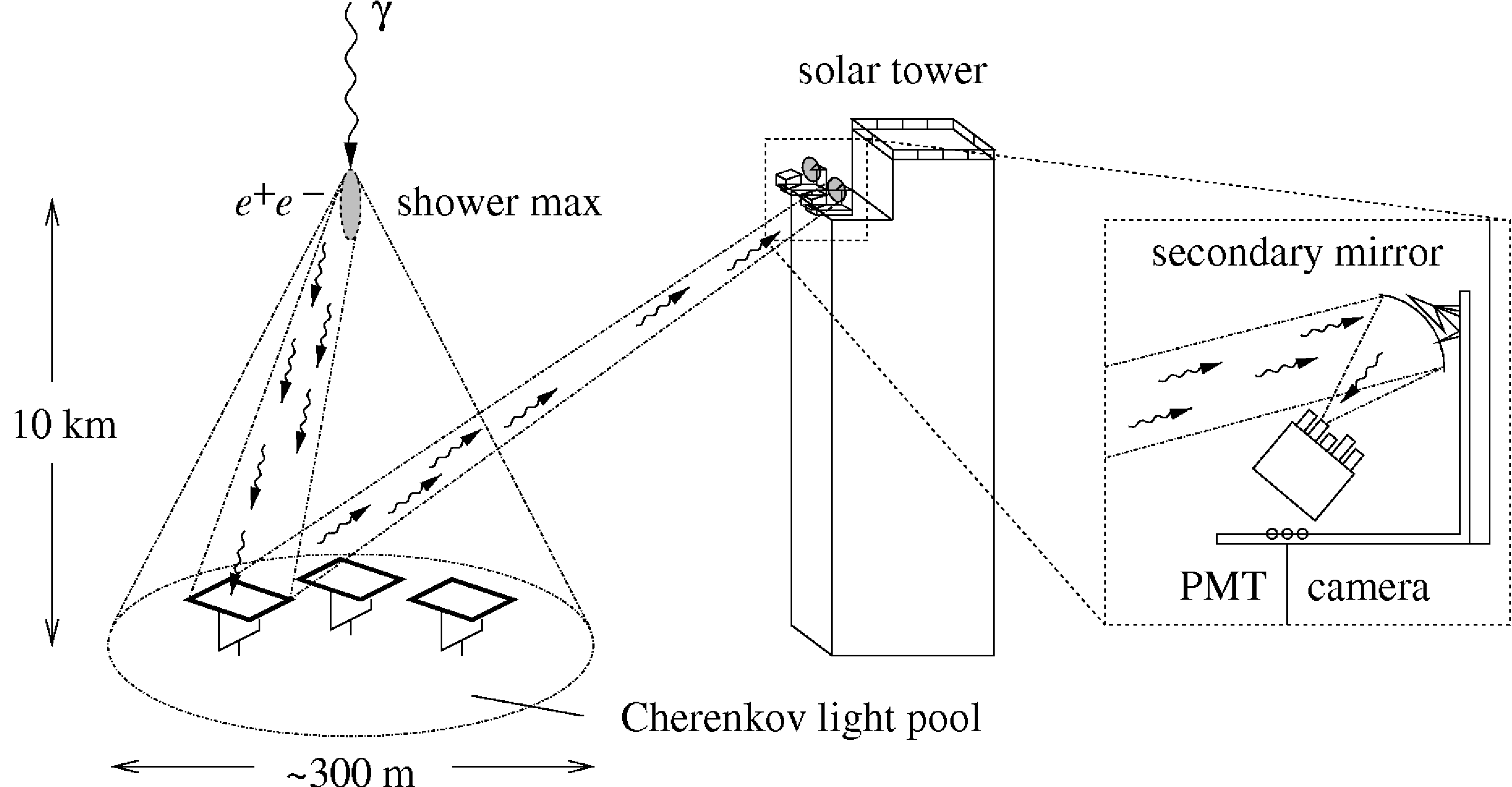}
\caption{A schematic representation of the wavefront sampling
technique. A gamma ray incident on the atmosphere induces an air
shower that illuminates an area approximately 300 m in diameter on the
ground with Cherenkov light. The light is reflected by heliostat
mirrors onto secondary mirrors and then focused onto PMTs mounted on
cameras under the secondaries. \label{fig: schematic}}
\end{figure}

STACEE is a ground-based atmospheric Cherenkov telescope that utilizes
the wavefront sampling technique \citep{smith06} to measure gamma rays
above about 100 GeV. Gamma rays and cosmic rays striking the Earth's
atmosphere induce extensive air showers, and Cherenkov light is one
product of the showers. The Cherenkov light pool typically illuminates
an area approximately 300 m in diameter on the ground. In the
wavefront sampling technique as employed by STACEE, the Cherenkov
light is focused by heliostat mirrors -- large reflectors built
originally for solar energy research -- onto secondary mirrors and
then onto photomultiplier tubes (PMTs). The digitized PMT signals form
the raw data product. There is one PMT viewing each heliostat, so the
technique samples the Cherenkov wavefront at the location of each
heliostat. Figure \ref{fig: schematic} illustrates this technique. A
key feature of all atmospheric Cherenkov detectors is that the
effective collecting area for high-energy gamma rays is much larger
than the physical size of the mirrors used to collect the Cherenkov
light. Wavefront sampling detectors like STACEE, which use large
heliostats as opposed to focusing dishes, have larger mirror areas and
can therefore work at lower gamma-ray energies where Cherenkov photon
densities on the ground are lower.

STACEE uses 64 of the 220 heliostats at the National Solar Thermal
Test Facility in Albuquerque, New Mexico. These 64 reflectors are
distributed fairly evenly over an area of \about 150 m \x 240 m. Each
heliostat is composed of a 5 \x 5 array of square mirror facets, and
each facet has a collecting area of 1.5 m$^2$. The light from a
heliostat is reflected onto one of five secondary mirrors. Three large
secondaries (\about 2 m in diameter) focus light onto one of three
cameras holding 16 PMTs each, and two smaller secondaries (\about 1.1
m in diameter) focus light onto one of two smaller cameras holding
eight PMTs each. STACEE also operates a custom-made atmospheric
monitoring system during observations. This system includes a weather
station and a photometer for measuring atmospheric transmission.

The signal from each PMT is fanned out to a discriminator and to an
eight-bit flash analog-to-digital converter (FADC). The FADCs
continuously digitize the 64 PMT signals with a sampling rate of 1
sample/ns; a typical Cherenkov pulse is several nanoseconds wide, so
these digitizers record the shape of each pulse. The pulse shape
information from the FADCs allows precise pulse times and sizes to be
extracted. Relative pulse timing is used to reconstruct the direction
of the gamma ray, and pulse sizes are used to reconstruct gamma-ray
energy. A set of dynamic delays ensures that the 64 pulses remain
synchronized as the source is followed across the sky.

The Cherenkov light from air showers arrives on the ground as a
``pancake'' a few nanoseconds thick, and over that brief interval it
is brighter than the night sky background. STACEE employs a two-level
trigger system and a tight timing coincidence requirement to pick out
Cherenkov-emitting air showers from the night sky glow. The trigger
system is designed as follows. First, the heliostats are grouped
geographically into eight clusters of eight heliostats each. The
cluster-level, or $L1$, trigger requires that the signals from five of
the eight heliostats within a cluster pass the discriminator
threshold, typically about four photoelectrons, within a timing window
of $8-24$ ns. Then, the array-level, or $L2$, trigger requires that
five of the eight clusters generate an $L1$ trigger within a 16-ns
window. When the $L2$ trigger condition is satisfied, the digitized
traces from all 64 PMTs are recorded.

The vast majority of the air showers detected by STACEE are initiated
by protons or light nuclei. In order to extract a gamma-ray signal
from data dominated by this cosmic-ray background, STACEE employs an
``on-off'' observing strategy in which each 28-minute on-source
observation is paired with an equal-exposure off-source
observation. The off-source observation is at the same declination as
the on-source observation and is 30 minutes (7.5 degrees) away in
right ascension, so it covers the same trajectory in local sky
coordinates. The number of cosmic-ray events in the off-source
observation is subtracted from the number of on-source events to find
the statistical excess of gamma-ray events from the direction of the
source. Many on-off pairs of observations for a given source taken
over several weeks are used together to establish a gamma-ray signal
from that source. The observations of Mrk 421 presented here were
taken on 18 nights from January 29 to May 15, 2004. The total
on-source exposure time was 20.5 hours, with an equal amount of
off-source exposure time.

\section{\label{sec: analysis}Analysis}

A set of criteria was applied to the data set to select and remove all
periods of time with unreliable data. First, data taken during a
hardware malfunction were removed. These malfunctions included
mispointings of one or more heliostats, trips of the high-voltage
power to the PMTs, or failures of the data acquisition system. Second,
the weather station data were checked for evidence of frost, which can
accumulate on the heliostats and seriously degrade their
reflectivities. A total of 5.7 hours of data was removed due to
hardware malfunctions and frost.

Data were also removed if marred by changes in background light
levels, \eg from clouds drifting through the field of view or from
lights from cars or airplanes passing near the experiment. The
observable quantities used to identify these time periods were the
$L1$ trigger rates. For every observing run, the $L1$ rates were
examined by eye and were required to vary smoothly during a run (\ie
no sudden spikes or drops). The stability of the $L1$ and $L2$ rates
were verified in the post-cut data. This step eliminated an additional
5.7 hours of on-source data. After all data quality cuts, the final
data set comprised 9.1 hours of on-source data taken on 16 nights.

Before the on- and off-source observations could be compared to
determine the statistical excess of gamma-ray events, an important
correction was made to account for the effects of the different
background light levels between the on- and off-source sky
fields. These differences are largely due to the particular stars in
those fields; a single bright star in one of the fields can radically
alter the light level. Low-energy cosmic-ray showers that are just
below the threshold required to initiate an event trigger can be
promoted above threshold by large positive fluctuations of the night
sky background. Likewise, showers that are just above threshold can be
demoted by large negative background fluctuations. If there were an
equal number of showers at all energies, then there would be the same
number of showers promoted above threshold as demoted below
threshold. However, because there are more low-energy showers than
high-energy ones (the cosmic-ray spectrum decreases with increasing
energy), the net effect of the night sky background fluctuations is to
promote subthreshold showers. This effect leads to a higher cosmic-ray
rate in observations with brighter sky fields. If the area of sky
covered by the on-source observation is brighter than that of the
off-source observation, as is the case for Mrk 421, then this effect
can imitate a true gamma-ray signal. 

A technique called ``padding'' \citep{scalzo04} has been developed to
equalize the backgroud light levels in a two-step process. The
technique relies on a pre-compiled library of FADC waveforms
corresponding to various low light levels measured under controlled
conditions. In the first step, background noise from this library is
added to ``pad'' the FADC waveforms of the observation with the dimmer
background. This step equalizes the RMS fluctuations, measured from
the 400 ns preceeding the pulse itself, between the on- and off-source
observations. In the second step, the trigger condition is reimposed
at a higher discriminator threshold. This is necessary because the
showers in the dimmer-background observation that \emph{would} have
triggered the detector if the light levels had been equal are absent
from the data stream. Imposing a software trigger identical to the
hardware trigger cannot regain these lost showers. Therefore, the
threshold is raised in software to the level where the
dimmer-background observations are not missing promoted showers
relative to the brighter-background observations. The technique is
fully described in \citet{scalzo04}, \citet{scalzophd04}, and
\citet{carson05}.

\section{\label{sec: int results}Integral Flux Results}

After applying the quality cuts and field brightness correction, there
are 2897 excess events in 511.3 minutes, for a photon count rate of
$5.7 \pm 0.9$ photons $\textrm{min}^{-1}$. The detection significance
is 5.9 standard deviations above the background. The mean significance
per observed pair is 1.4 $\sigma$ in 28 minutes of on-source exposure
time.

\begin{figure}
\epsscale{1.0}
\plotone{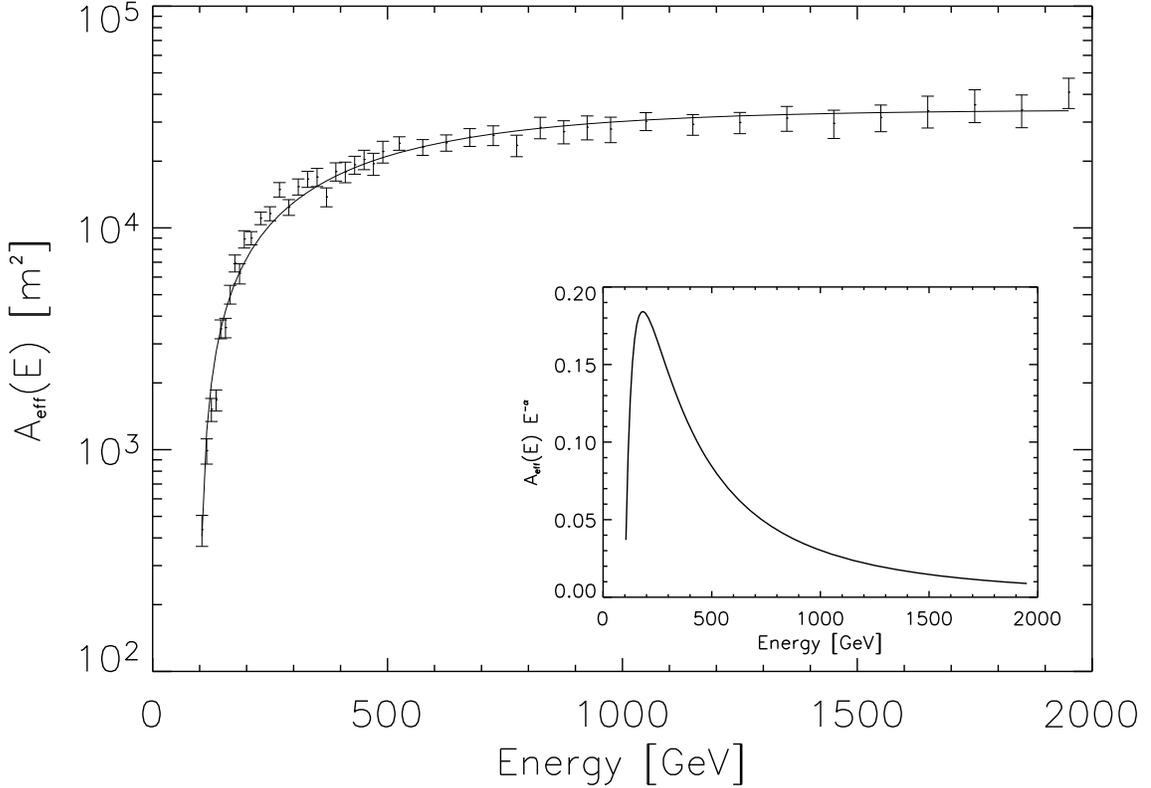}
\caption{\emph{main figure:} The effective area, the overall
collecting area of the experiment, for the Mrk 421 data set. The data
points represent an exposure-weighted average over simulated
observations at five different zenith angles, and the solid line is a
fit to the form $A_0 \left(1-e^{\frac{-(E-E_{min})}{E_r}}\right)$ where
$A_0$, $E_{min}$, and $E_r$ are constants. \emph{inset:} The
(unnormalized) differential trigger rate as a function of energy for a
spectral index $\alpha = 2.0$. The energy threshold is defined as the
peak of this curve (180 GeV). \label{fig: trig rate}}
\end{figure}

The detector response is characterized by the effective area
$A_{eff}$, defined as the overall gamma-ray collecting area of the
experiment, as opposed to the physical area of the heliostat
mirrors. It is determined from Monte Carlo simulations of the air
shower physics and the detector performance, as described in
\citet{scalzo04} and \citet{bramel05}. The effective area for the Mrk
421 data set is shown in Figure \ref{fig: trig rate}. The data points
reflect simulations of observations at five different zenith angles,
exposure-weighted to reflect the composition of the real data set. The
solid line is a fit to the data assuming $A_{eff} = A_0 \left(1 -
e^{\frac{-(E-E_{min})}{E_r}}\right)$, where $A_0$, $E_{min}$, and
$E_r$ are constants.

The differential trigger rate $\frac{dR}{dE}$ (photons s$^{-1}$
GeV$^{-1}$) is $A_{eff}$ times the differential photon flux
$\frac{dN}{dE}$, expressed as a power law:

\begin{displaymath}
\frac{dN}{dE} = N_0 \,\, \left(\frac{E}{E_0}\right)^{-\alpha} \quad .
\end{displaymath}

\noindent Assuming a spectral index $\alpha = 2.0$, the curve of
$A_{eff} E^{-\alpha}$ for the Mrk 421 data set, which is proportional
to the differential trigger rate, is shown in the inset to Figure
\ref{fig: trig rate}.

Following the usual convention in VHE astronomy, we define the energy
threshold as the energy of the peak of the differential rate
curve. From Figure \ref{fig: trig rate}, the peak is at $E_{th} = 180$
GeV. It is clear from the figure that, like all VHE instruments,
STACEE has significant sensitivity to energies below $E_{th}$, as
defined in this way. The systematic error on the STACEE energy scale
is estimated to be \about 20\% \citep{scalzophd04}. Assuming $\alpha =
2.0$, the integral flux of VHE gamma rays from Mrk 421 measured by
STACEE is $\Phi (E > 180 ~ \textrm{GeV} = (4.7 \pm 0.8_{stat}) \times
10^{-10} \,\, \textrm{photons} \,\, \textrm{cm}^{-2} \,\,
\textrm{s}^{-1}$.

\section{\label{sec: spec results}Spectral Analysis and Results}

We have developed a technique for reconstructing gamma-ray energies
from the charges recorded on each of the 64 PMTs that constitute the
STACEE camera. The method relies on two physical properties of
gamma-ray showers: 1) the intensity of the Cherenkov light on the
ground is directly proportional to the energy of the initiating gamma
ray and 2) the light distribution on the ground is approximately
uniform out to the edges of the shower. These properties, along with
an independent estimate of the shower core location
\citep{scalzophd04}, allow us to convert a PMT charge to a number of
photons arriving at the corresponding heliostat. This conversion
relies on simulations of the STACEE electronics and optics that have
been thoroughly checked against calibration data \citep{scalzophd04,
carson05}. The reconstructed lateral photon distribution is then
converted to a gamma-ray energy using standard simulations of air
showers initiated by gamma rays. The energy reconstruction method
successfully reconstructs the energies of simulated gamma rays between
130 GeV and \about 2 TeV with an energy resolution of $20-30\%$;
systematic errors in the energy estimate are less than $10\%$. The
details of the technique are described in \citet{carson05}. It was
applied to STACEE data taken on the Crab Nebula, the standard
calibration source for VHE astronomy, and the results are in good
agreement with previous measurements
\citep{hillas98,aharonian04,wagner05}.

The energy reconstruction method was also applied to the 2004 Mrk 421
data. The reconstructed energies were sorted into six energy bins:
$130-150$ GeV, $150-200$ GeV, $200-300$ GeV, $300-500$ GeV, $500-1000$
GeV, and $>1$ TeV. The bin choices were motivated by two
considerations: 1) ensuring that the effective area did not change
substantially over the bin, and 2) having a similar number of entries
in each bin. For each energy bin, the number of deadtime-corrected
entries in the off-source observations was subtracted from the number
of deadtime-corrected entries in the on-source observations, and the
gamma-ray excess in each energy bin was divided by the livetime for
the observations and the time-averaged effective area at that
energy. The resulting differential spectrum $\frac{dN}{dE}$ (photons
cm$^{-2}$ s$^{-1}$ GeV$^{-1}$) is shown in Figure \ref{fig: spectrum}.
Fitting the spectrum with the power law from Section \ref{sec: int
results} with $E_0 = 100$ GeV, we find

\begin{displaymath}
N_0 = (9.6 \pm 3.5_{stat} \, {^{+3.1} _{-2.0}}_{sys}) \times 10^{-12}
\quad \textrm{photons} \, \, \textrm{cm}^{-2} \, \, \textrm{s}^{-1} \,\, 
\textrm{GeV}^{-1}
\end{displaymath}

\noindent and

\begin{displaymath}
\alpha = 2.1 \pm 0.2_{stat} \, {^{+0.2} _{-0.1}}_{sys}  \quad .
\end{displaymath}

\noindent In the inset to Figure \ref{fig: spectrum}, the Mrk 421 SED
is plotted as $\nu F_{\nu} = E^2 \frac{dN}{dE}$ (ergs cm$^{-2}$
s$^{-1}$).

\section{\label{sec: discussion}Discussion}

\begin{figure}
\epsscale{1.0} \plotone{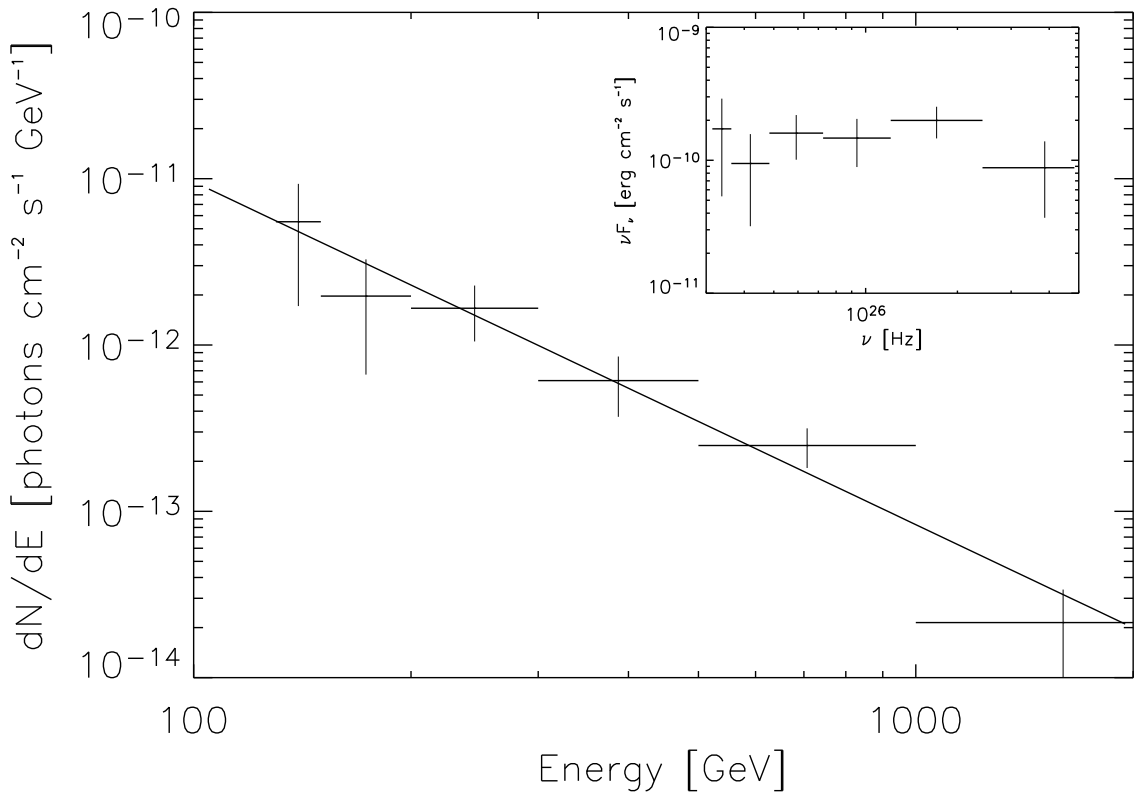}
\caption{\emph{main figure:} The Mrk 421 spectrum between 130 GeV and
2 TeV, as measured by STACEE in early 2004. The spectral index is
$\alpha = 2.1 \pm {0.2_{stat}} \, {^{+0.2} _{-0.1}}_{sys}$. The
vertical error bars represent the statistical errors in both the
signal and the effective area. The horizontal bars represent the
widths of the energy bins. \emph{inset:} The same spectrum plotted
using the standard convention, as $\nu F_{\nu} = E^2 \frac{dN}{dE}$
(ergs cm$^{-2}$ s$^{-1}$) vs. frequency $\nu$. \label{fig: spectrum}}
\end{figure}

As shown in the inset to Figure \ref{fig: spectrum}, the Mrk 421 SED
is flat in the STACEE band. The STACEE measurement suggests that the
high-energy peak of the SED of Mrk 421 was at or above 130 GeV (log
$\nu = 25.5$) in early 2004. This is a slightly higher energy for the
peak than predicted by most SSC models, such as the time-dependent SSC
model of \citet{krawczynski01}, which predicts a high-energy peak of
between 10 and 100 GeV. A shift of the high-energy peak above 100 GeV
is sometimes invoked in modeling of bright flares, however; for
instance, \citet{maraschi99} suggested that the peak shifted to
several hundred GeV during a flare in April 1998.

The STACEE observations of Mrk 421 were part of an extensive
multiwavelength campaign; during the first five months of 2004, the
source was monitored by radio, optical, X-ray (\emph{RXTE}), and TeV
(Whipple) telescopes simultaneously. These multiwavelength
measurements were presented in Figure 12 of \citet{blazejowski05} and
are summarized here. The data were divided into ``medium'' and
``high'' states of activity based on their X-ray fluxes. It turned out
that this division separated the data into two time periods; the
medium state covers the period from January 27 to March 26 2004, and
the high state covers April $16-20$ 2004. The medium- and high-state
data were fit to one- and two-zone SSC models, respectively, where a
zone is a spherical emitting region characterized by a single magnetic
field and electron population. These models \citep{krawczynski04} are
snapshots of the SSC radiation emitted by a steady-state electron
population, and do not evolve the electron spectrum
self-consistently. Power law fits were performed on the Whipple data
assuming a high-energy cutoff of 4.96 TeV; these fits returned
spectral indices of $2.40 \pm 0.18$ and $2.11 \pm 0.14$ for the
medium- and high-state data, respectively. The STACEE data set covers
about 40\% of the nights represented in the multiwavelength SED
(Figure 12 of \citet{blazejowski05}), and \about 90\% of the STACEE
data set was taken during these nights. Approximately 25\% of the
STACEE data were from the TeV flaring state in April.

\begin{figure}
\epsscale{1.0}
\plotone{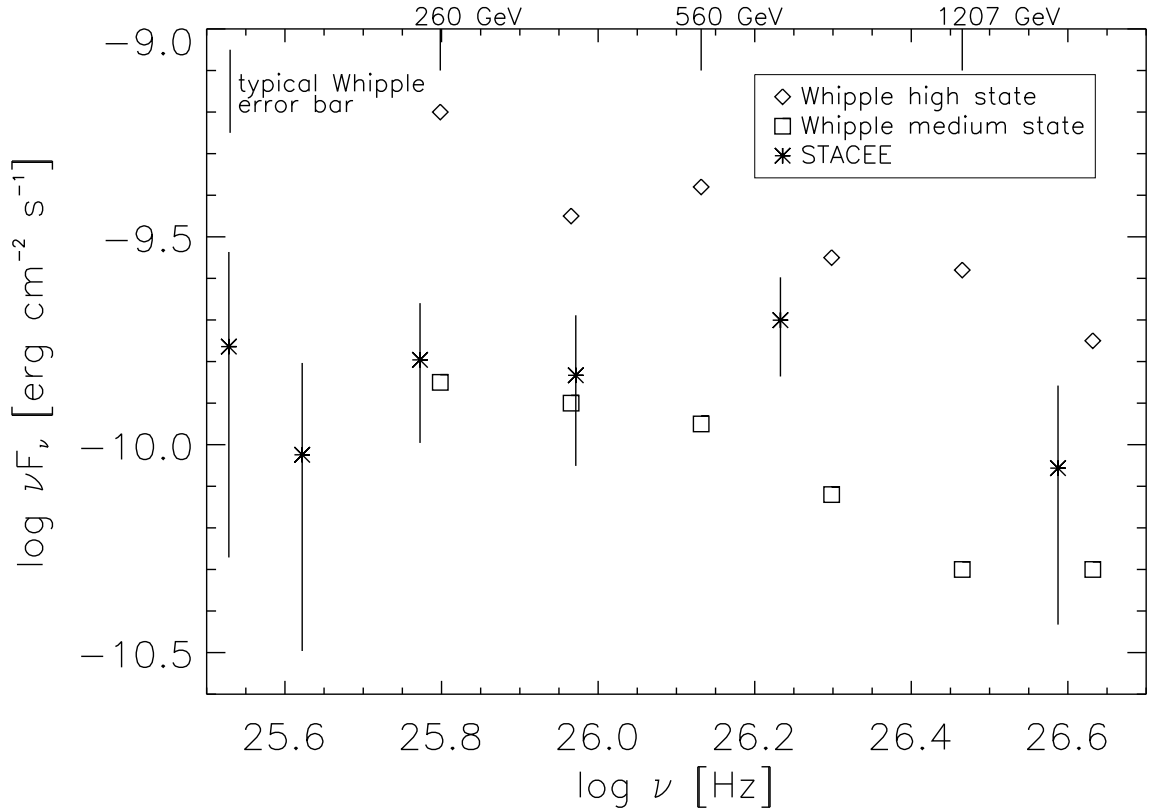}
\caption{The spectral energy distributions of Mrk 421 from Whipple
(open symbols) and STACEE (stars). The former are divided into medium
(squares) and high (diamonds) states of activity based on their X-ray
flux, as in \citet{blazejowski05}. A typical Whipple error bar is
shown in the upper left corner. \label{fig: tevcomp}}
\end{figure}

The STACEE spectral index is consistent with the high-state spectral
index determined by Whipple and marginally consistent with the
medium-state spectral index. However, it is naive to expect the
spectrum to follow a simple power law over such a large range in
energy, and the most useful comparison of the two data sets is
restricted to the energy range where they overlap, as shown in Figure
\ref{fig: tevcomp}. The open symbols are the Whipple data from the
high (diamonds) and medium (squares) states. The STACEE spectrum is
shown as asterisks. We note that the first STACEE energy bin is at 140
GeV, 120 GeV lower than the first Whipple energy bin. Figure \ref{fig:
tevcomp} shows that the STACEE and Whipple measurements are in good
agreement, especially considering that no cross-calibration was
performed on the data from the two instruments and that the errors
shown are statistical only. We expect the STACEE flux levels to
coincide more closely with the Whipple measurements in the medium
state, as seen, since the majority of the STACEE data were taken
before the April flare. The STACEE data show that the spectrum remains
approximately flat down to an energy of 130 GeV.

The most consistent interpretation of the combined STACEE and Whipple
data is that the peak of the SED is around $130-500$ GeV (log $\nu
\sim 25.5 - 26.1$). This is a higher peak energy than expected from
past modeling for a medium flux state, including the SSC model shown
in \citet{blazejowski05}. We note, however, that the
\citet{blazejowski05} medium-state model overpredicts the Whipple data
at the lowest energies, and a peak in the range indicated above is a
better reflection of the data than the one-zone model peak. A
detection of the peak indicates that we are measuring most of the
energy released in inverse-Compton cooling, and constrains the
inverse-Compton energy much more than a measurement of the falling
edge of the SED alone. Going a step further and combining the VHE data
with the X-ray data near the synchrotron peak then measures the total
energy budget of Markarian 421 and the relative contributions of
synchrotron and inverse-Compton cooling. During the period of these
observations, the combined STACEE and X-ray data suggest that the
synchrotron peak was a factor of 2-3 higher than the inverse-Compton
peak, suggesting that synchrotron cooling dominates. Because these
data represent a time average of the jet radiation, we suggest further
modeling that evolves the electron population as it cools via
synchrotron radiation.

During the preparation of this manuscript we learned of a measurement
of the Mrk 421 energy spectrum by the MAGIC collaboration
\citep{albert06}. The energy range covered by MAGIC includes much of
that explored by STACEE and they also report a flattening of the
energy spectrum at the low end. Their data are from a different epoch
(November 2004 to April 2005), so a direct comparison of data sets
from this highly episodic source is not easily made. However, it is
clear that many such observations will help with the detailed
understanding of Mrk421.

In summary, we have measured a spectrum of Markarian 421 above 130 GeV
in six energy bins. The STACEE data suggest that we are detecting the
high-energy peak of the SED for the first time above 100 GeV and that
the peak is around $130-500$ GeV. The direct detection of the
high-energy peak can strongly constrain future modeling.

\acknowledgments

We are grateful to the staff at the National Solar Thermal Test
Facility, who continue to support our science with enthusiasm and
professionalism. This work was supported in part by the National
Science Foundation, the Natural Sciences and Engineering Research
Council of Canada (NSERC), Fonds
Qu$\acute{\textrm{e}}$b$\acute{\textrm{e}}$cois de la Recherche sur la
Nature et les Technologies (FQRNT), the Research Corporation, and the
University of California at Los Angeles.

\clearpage

\clearpage

\clearpage

\clearpage

\clearpage

\end{document}